\documentclass[10pt,conference,a4paper]{IEEEtran}

\usepackage{cite}
\usepackage{amsmath}
\usepackage{amssymb}
\usepackage{graphicx}
\usepackage{textcomp}
\usepackage{xcolor}
\usepackage{booktabs}
\usepackage{url}
\usepackage{hyperref}
\usepackage{tikz}
\usetikzlibrary{positioning, arrows.meta, fit, backgrounds, shapes.geometric}
\usepackage{multirow}

\begin{document}

\title{Beyond Document Retrieval: Architectural Challenges When LLM Agents Query Structured Enterprise Data}

\author{\IEEEauthorblockN{Sheikh Nazib Ahmed}
\IEEEauthorblockA{University of Texas at Arlington \\
Arlington, TX, USA \\
sxa5256@mavs.uta.edu}}

\maketitle

\begin{abstract}
Retrieval-augmented generation (RAG) has become a common architecture for connecting large language models to enterprise knowledge. Most RAG systems retrieve unstructured documents---PDFs, wiki pages, support tickets---and feed them to an LLM for summarization or question answering. A growing class of enterprise agents, however, must query \textit{structured data}: relational databases, data warehouses, and analytics APIs where the answer is a computed result, not a retrieved passage. Structured-data querying forces decisions that a document-RAG pipeline never has to make. We group them into seven dimensions: retrieval semantics, authorization, intent recognition, entity resolution, evaluation, failure modes, and latency. For each dimension, we characterize the baseline assumption, explain its limitation for structured data, and describe a generic architectural pattern. As supporting evidence, a controlled synthetic study shows that a staged agent built on this framework eliminates the authorization violations of a direct translate-and-execute baseline under controlled conditions. The primary result is a design-oriented framework, an evaluation protocol, and a set of open problems for governed structured-data agents.
\end{abstract}


\section{Introduction}
\label{sec:intro}

When a team wants an LLM grounded in their own data, they usually reach for retrieval-augmented generation (RAG)~\cite{lewis2020rag}. The recipe is familiar by now. Documents are chunked, embedded, and stored in a vector index, and at query time the system pulls the top-$k$ nearest chunks and feeds them to the model as context. Most of the RAG literature then tunes one stage or another, from chunking~\cite{chen2024benchmarking} and embedding models~\cite{muennighoff2022mteb} to retrieval~\cite{karpukhin2020dpr}, reranking~\cite{nogueira2020passage}, and prompt construction~\cite{gao2023ragsurvey}.

This architecture assumes the data source is a corpus of documents. The user asks a question, the system finds the most relevant passages, and the LLM synthesizes an answer from those passages. But a growing number of enterprise AI agents do not retrieve documents at all. They query \textit{structured data}---relational databases, data warehouses, analytics APIs---where the answer is a computed result: a count, a filtered table, an aggregated metric.

Consider two questions posed to an enterprise AI agent:
\begin{itemize}
    \item \textit{``What is our company's return policy?''} --- The agent retrieves relevant policy documents and synthesizes an answer. This is document RAG.
    \item \textit{``How many open orders does Acme Corp have?''} --- No document contains this answer. The agent must translate the question into a SQL query, execute it against a database, and summarize the numeric result. This is a structured-data agent.
\end{itemize}

The second question requires a different sequence of responsibilities. The agent must identify the target domain, resolve the customer entity to a canonical identifier, verify customer and domain access, select the appropriate data source, generate a schema-compliant query or API request, execute it, apply any required result restrictions, and summarize the returned data. Some of these responsibilities can occur in other AI systems, but they are not optional implementation details when the answer depends on live, access-controlled structured data.

A document-RAG template can therefore be a poor starting abstraction for structured-data agents if its retrieval, authorization, and intent assumptions are carried over without redesign. A governed structured-data design must make source routing, policy enforcement, entity binding, and query validation explicit rather than leaving them implicit in document similarity or a single prompt. These requirements motivate a comparison of the two architectural patterns, but we do not claim that every RAG system or every structured-data agent follows the same design.

This paper examines what changes architecturally when the data source shifts from unstructured documents to structured data. We identify seven dimensions where a minimal document-RAG baseline does not expose requirements that become explicit in structured-data agents. We examine each dimension through representative enterprise scenarios and a generic reference architecture. The analysis focuses on reusable responsibilities and control dependencies rather than claiming that one organization's design is universal.

Our contributions are:
\begin{enumerate}
    \item A design-oriented comparison of minimal document-RAG and structured-data-agent designs across seven architectural dimensions.
    \item A reference architecture that maps structured-data requirements to explicit routing, authorization, resolution, validation, execution, and persistence stages.
    \item A failure-mode taxonomy that organizes five recurring risks that are particularly salient when LLMs generate and execute structured-data queries.
    \item Identification of four open problems that current enterprise structured-data agents do not fully solve.
    \item A controlled synthetic study showing that the staged design improves outcome accuracy and eliminates authorization violations relative to a direct translate-and-execute baseline.
\end{enumerate}

\section{Background}
\label{sec:background}

\subsection{Document-RAG Architecture}

The canonical RAG architecture~\cite{lewis2020rag} operates in two phases. During \textit{indexing}, documents are split into chunks, each chunk is converted to a dense vector embedding, and the embeddings are stored in a vector index. During \textit{inference}, the user's query is embedded, the top-$k$ nearest chunks are retrieved, and the retrieved chunks are concatenated with the query as context for the LLM. The LLM generates a response grounded in the retrieved passages.

Extensions to this architecture include multi-step retrieval~\cite{gao2023ragsurvey}, hybrid search (combining dense and sparse retrieval), query rewriting, and agentic RAG where an LLM decides which retrieval actions to take. Despite these extensions, the core assumption persists: the data source is a collection of text passages, and the system's job is to find the right passages and let the LLM read them.

\textbf{What RAG optimizes for.} The extensive RAG optimization literature focuses on three quality dimensions: (1) retrieval relevance---ensuring the right chunks are found, (2) context construction---arranging the retrieved chunks into an effective prompt, and (3) generation faithfulness---ensuring the LLM's response is grounded in the retrieved context. These three dimensions define the quality model for document RAG. As we show in Section~\ref{sec:dimensions}, structured-data agents require a different quality model entirely: query correctness, schema fidelity, authorization compliance, entity resolution accuracy, and summarization faithfulness replace the three document-RAG dimensions.

\subsection{Structured-Data Agents}

Structured-data agents translate natural-language questions into formal queries---typically SQL---against relational data sources. This is the text-to-SQL problem~\cite{yu2018spider}, now extended with LLMs as the translation engine. In an enterprise setting, the agent must handle multiple data sources, enforce access control, resolve entity references, and present results in natural language.

The key difference from document RAG is that the LLM does not read a passage and synthesize an answer. Instead, it generates executable code (a query), the system executes that code against a database, and the raw results are returned for summarization. The LLM operates as a \textit{translator and summarizer}, not as a \textit{reader and synthesizer}.

\subsection{Scope of the Design Study}

We consider a governed structured-data agent that serves multiple access personas and answers questions over several operational data sources. The agent may support direct lookups, multi-source comparisons, and sequential investigations. A generic design includes question interpretation, policy checks, entity binding, schema and source planning, answerability validation, governed query execution, response explanation, and audit logging. These are architectural capabilities rather than a prescribed implementation order; a deployment may combine or split them according to its data services and governance model.

Users interact with the agent through a natural-language interface and receive a result together with an explanation of the answer. A request may terminate early when its scope is unsupported, access is denied, an entity is ambiguous, a source is unavailable, or the available context is insufficient for a safe query. The reference architecture treats these outcomes as explicit control-flow states rather than as silent retrieval failures.

\subsection{Design Analysis Method}
We use a structured design analysis rather than a deployment report. First, we compare the assumptions of a minimal document-RAG baseline with the responsibilities exposed by structured-data querying. Next, we organize those responsibilities into seven dimensions and trace their interactions through representative scenarios. Finally, we derive a reference architecture, a failure taxonomy, and a stage-aware evaluation protocol. The paper reports no organization-specific prompts, schemas, endpoints, logs, deployment measurements, or proprietary evaluation results. The examples are intentionally abstract so that the analysis can be discussed independently of any particular product or employer.

\section{Seven Architectural Dimensions}
\label{sec:dimensions}

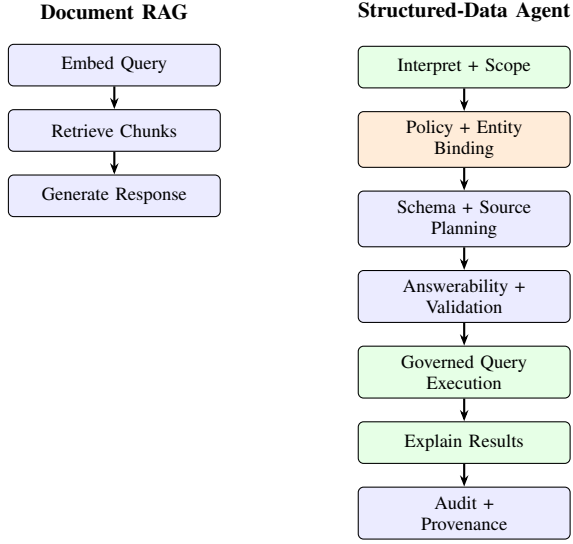
\begin{figure}[t]
\centering
\begin{tikzpicture}[
    node distance=0.35cm,
    block/.style={rectangle, draw, fill=blue!8, minimum height=0.55cm, minimum width=2.8cm, align=center, font=\scriptsize, rounded corners=2pt},
    arr/.style={-{Stealth[length=1.5mm]}, thick},
    lbl/.style={font=\footnotesize\bfseries},
]

\node[lbl] (dlbl) {Document RAG};
\node[block, below=0.2cm of dlbl] (d1) {Embed Query};
\node[block, below=0.3cm of d1] (d2) {Retrieve Chunks};
\node[block, below=0.3cm of d2] (d3) {Generate Response};
\draw[arr] (d1) -- (d2);
\draw[arr] (d2) -- (d3);

\node[lbl, right=2.0cm of dlbl] (slbl) {Structured-Data Agent};
\node[block, below=0.2cm of slbl, fill=green!10] (s1) {Interpret + Scope};
\node[block, below=0.3cm of s1, fill=orange!15] (s2) {Policy + Entity\\Binding};
\node[block, below=0.3cm of s2] (s3) {Schema + Source\\Planning};
\node[block, below=0.3cm of s3] (s4) {Answerability +\\Validation};
\node[block, below=0.3cm of s4, fill=green!10] (s5) {Governed Query\\Execution};
\node[block, below=0.3cm of s5, fill=green!10] (s6) {Explain Results};
\node[block, below=0.3cm of s6] (s7) {Audit +\\Provenance};
\draw[arr] (s1) -- (s2);
\draw[arr] (s2) -- (s3);
\draw[arr] (s3) -- (s4);
\draw[arr] (s4) -- (s5);
\draw[arr] (s5) -- (s6);
\draw[arr] (s6) -- (s7);

\end{tikzpicture}
\caption{Condensed pipeline comparison. Document RAG is shown as embed, retrieve, and generate. A generic structured-data design makes scope, policy, entity binding, source planning, validation, execution, and response handling explicit; deployments may combine or reorder these capabilities.}
\label{fig:pipelines}
\end{figure}

We identify seven dimensions where the shift from document retrieval to structured-data querying requires different architectural patterns. Figure~\ref{fig:pipelines} contrasts the two pipelines at a high level, and Table~\ref{tab:comparison} provides a detailed comparison across all seven dimensions.

\begin{table*}[t]
\centering
\caption{Architectural comparison across seven dimensions. Each row describes a document-RAG assumption and how structured-data agents diverge.}
\label{tab:comparison}
\begin{tabular}{p{2.4cm}p{5.5cm}p{5.5cm}}
\toprule
\textbf{Dimension} & \textbf{Document RAG} & \textbf{Structured-Data Agent} \\
\midrule
Retrieval semantics & Semantic similarity over text chunks & Schema-aware source planning and query construction \\
\addlinespace
Authorization & Document-level ACLs; filter at retrieval & Entity-, scope-, or tenant-level policy checks before execution \\
\addlinespace
Intent recognition & Implicit in embedding similarity & Explicit task, scope, and entity state for downstream planning \\
\addlinespace
Entity resolution & Passage ranking leaves ambiguity to generation & Explicit binding of references to canonical source identifiers \\
\addlinespace
Evaluation & Retrieval quality (recall@k, MRR) + generation quality (faithfulness) & Query, policy, entity, result, and explanation correctness \\
\addlinespace
Failure modes & Retrieval misses, hallucination from irrelevant context & Schema hallucination, wrong joins, identifier confusion, partial authorization \\
\addlinespace
Latency profile & Embedding lookup + LLM call & Model calls + policy and metadata services + query execution \\
\bottomrule
\end{tabular}
\end{table*}

\subsection{Dimension 1: Retrieval Semantics}

Document-RAG retrieval works by semantic similarity: embed the query, find the nearest chunks in vector space, feed them to the LLM. Failures show up as irrelevant or missing chunks, and the entire optimization literature focuses on making this similarity search better.

Structured-data agents have no text to retrieve. The system must translate the user's question into a formal query that the database can execute. This translation requires understanding the database schema: which tables exist, what columns they contain, what join paths connect them, and what filters to apply. The LLM must produce syntactically and semantically correct SQL, not find a similar passage.

This changes the retrieval pipeline entirely. Instead of a vector database, the system needs a \textit{schema registry}---a catalog of available tables, their columns, descriptions, and relationships. Instead of embedding similarity, the system needs \textit{schema-aware routing}: determining which table(s) can answer the user's question based on column coverage and domain alignment.

\textbf{Scenario.} A user asks ``what is the contract end date for Acme Corp?'' In document RAG, the system retrieves paragraphs mentioning Acme Corp and contracts. In a structured-data agent, the system must identify that the \texttt{contracts} table has a \texttt{contract\_end\_date} column, that the customer can be identified via a customer key, and generate a query joining the contracts table with the customer lookup. No text is retrieved; a query is constructed.

\textbf{Multi-source complexity.} A multi-source agent must account for related tables with different scopes, identifiers, and column sets. A schema registry and source-planning component can select candidate sources and expose valid join paths before query construction. Table selection is therefore a schema-and-scope routing problem rather than ordinary passage retrieval. The selected data surface also determines which query and policy checks are required later in the workflow.

\subsection{Dimension 2: Authorization Granularity}

Document-RAG authorization commonly filters documents using document-level access-control metadata before the retrieved context reaches the LLM~\cite{arbiter2025, permaware2025}. Structured-data agents must bind authorization decisions to the entities and domains named by the query.

A governed structured-data design may separate policy checks into several scopes:

\begin{itemize}
    \item \textbf{Entity scope:} Can this user access the entity named or implied by the question?
    \item \textbf{Data scope:} Can this user's role or tenant access the requested business area and source?
    \item \textbf{Field scope:} Which attributes, if any, require masking or suppression after source selection?
\end{itemize}

These checks should be enforced outside the language model and treated as separate boundaries. They are query-dependent because the system must first determine which entity, scope, and source the question refers to. Independent policy lookups may run concurrently with interpretation, but a policy result must constrain candidate selection before an ambiguous entity is presented to a user.

\textbf{Partial authorization.} A cross-source question may include both permitted and non-permitted data scopes. A governed agent must decide whether to reject the request, return an authorized subset with an explicit explanation, or ask the user to narrow the scope. This is a useful architectural distinction from a simple document filter.

\textbf{Authorization timing.} Policy checks should occur before query execution and, where candidate entities are displayed, before candidate selection. Scope-dependent checks may follow task interpretation because the system needs the resolved data area. The important invariant is that untrusted model output cannot expand the authorized data set.

\subsection{Dimension 3: Intent Recognition}

A document retriever may provide topical routing implicitly through similarity, but a structured-data agent must produce explicit task and scope state because downstream registry and policy decisions depend on it. The interpretation component may identify one or more data areas, distinguish a direct lookup from a comparison or investigation, and record uncertainty for later handling.

A generic catalogue contains descriptions of available business areas and their source coverage. A fallback handles questions that are irrelevant, too vague, or outside the catalogue. Trusted application context may reduce the amount of interpretation required, but it should not bypass policy enforcement. Independent interpretation and policy lookups can run concurrently when neither depends on the other's result.

\textbf{Multi-source intent.} A question can require multiple data areas without naming them explicitly. For example, a service-status question may require operational, inventory, and support sources. The intent component therefore returns a set of candidate scopes and a task type rather than a single retrieval label. This explicit state enables source planning to choose between direct lookup, comparison, and sequential investigation.

\subsection{Dimension 4: Entity Resolution}

Retrieval ranking can leave ambiguity for the LLM to interpret, but a structured-data query must bind an entity reference to a canonical identifier before access checks and query execution. If multiple customer records match a name, the system cannot safely select an arbitrary result.

A governed entity-binding component should apply the user's policy context before presenting candidate records. It may automatically accept a unique authorized match, request clarification when several candidates remain, or terminate when no authorized candidate exists. The selected canonical reference is then passed to source planning and query construction; execution should not proceed while the identity decision is unresolved.

\textbf{Identifier diversity.} Different sources may require different identifiers even when they represent the same real-world entity. A registry can map a canonical reference to source-specific identifiers and valid translation paths. This mapping is a data-integration responsibility, not a prompt-only instruction, because the wrong identifier can produce a syntactically valid but semantically empty or incorrect query.

\textbf{Human-in-the-loop protocol.} Entity ambiguity creates a first-class control-flow state: pause, present sanitized candidate labels, accept a selection, and resume with the chosen canonical identifier. This is materially different from returning the nearest document passage, because the system must not execute the query until the identity decision is resolved.

\subsection{Dimension 5: Evaluation}

Evaluating a document-RAG system is straightforward: measure retrieval quality (recall@k, MRR, NDCG) and generation quality (faithfulness, relevance, answer correctness~\cite{chen2024benchmarking}). Two stages, well-understood metrics.

Structured-data agents are harder to evaluate because the pipeline has more stages and different failure modes at each one:

\begin{itemize}
    \item \textbf{Query correctness:} Does the generated SQL/API call return the correct result? (execution accuracy~\cite{li2024bird})
    \item \textbf{Schema fidelity:} Does the query reference only real tables and columns? (no hallucinated schema elements)
    \item \textbf{Authorization compliance:} Did the system correctly enforce all authorization layers?
    \item \textbf{Entity resolution accuracy:} Did the system resolve the correct entity?
    \item \textbf{Summarization quality:} Does the natural-language response accurately reflect the query results?
\end{itemize}

No single metric captures end-to-end quality. Evaluation must be stage-aware, measuring each pipeline stage independently and in combination.

\textbf{Ground truth construction.} In document RAG, ground truth is a set of question-answer pairs where the answer can be traced to specific source passages. In structured-data agents, ground truth requires correct SQL queries for each question---and the ``correct'' query may not be unique. Two queries with different syntax can return identical results (e.g., using \texttt{WHERE} vs.\ \texttt{HAVING} for the same filter). Execution accuracy (comparing query results rather than query text) addresses this, but introduces the overhead of maintaining a test database with known data.

\textbf{Multi-stage failure attribution.} When a structured-data agent returns a wrong answer, the error could originate in task interpretation, entity binding, source planning, query generation, policy enforcement, or explanation. Attributing the failure to the correct stage requires recording intermediate decisions and inputs. A stage-aware evaluation harness should score these outputs separately and support controlled tests with known upstream state.

\begin{table}[t]
\centering
\caption{Stage-aware evaluation protocol for governed structured-data agents. The entries define test outputs rather than results from a particular deployment.}
\label{tab:evaluation}
\scriptsize
\resizebox{\columnwidth}{!}{%
\begin{tabular}{p{1.7cm}p{2.0cm}p{2.4cm}}
\toprule
\textbf{Stage} & \textbf{Output} & \textbf{Suggested measure} \\
\midrule
Interpretation & Task and data scope & Top-$k$ accuracy; coverage \\
Entity binding & Canonical reference & Resolution accuracy; abstention rate \\
Policy gate & Allowed data set & Compliance; leakage rate \\
Source planning & Tables and join paths & Set F1; execution validity \\
Query and answer & Query result and explanation & Execution accuracy; faithfulness \\
\bottomrule
\end{tabular}
}
\end{table}

This protocol separates architectural behavior from end-to-end answer quality. It also makes it possible to compare systems without requiring one canonical query string: equivalent queries can be judged by their results, policy behavior, and provenance.

\subsection{Controlled Synthetic Study}
\label{sec:synthstudy}

To test whether the staged design changes outcomes, we ran a controlled experiment on a synthetic multi-source benchmark. Two agents answer an identical set of 21 questions spanning four access roles and two independent data sources (an operations store and a compliance store). The \emph{baseline} agent translates each question into a query and executes it directly. The \emph{staged} agent adds the reference-architecture stages: a policy gate, policy-aware entity binding, schema and join validation, and abstention when context is insufficient. Both agents receive the same parsed intent, so the only independent variable is the architecture. All data is synthetic and seeded, and ground truth is computed by an independent reference implementation that derives each correct outcome from the policy specification and hand-verified reference queries rather than from either agent.

Table~\ref{tab:results} reports the comparison. The staged agent reaches 0.95 outcome accuracy against 0.43 for the baseline, and removes the seven authorization violations and eight confidently wrong answers that the baseline produces (leakage rate 0.00 vs.\ 0.33). The gains come from behavior the baseline cannot express: rejecting out-of-scope requests, abstaining on ambiguous entities and unsupported metrics, and refusing invalid cross-source joins. Answer correctness on the cases both agents choose to answer is identical (0.90), which indicates the improvement is due to governance and validation rather than a stronger query generator. This is a small synthetic study rather than a production measurement; it shows that the architectural stages have a measurable effect under controlled conditions.

\begin{table}[t]
\centering
\caption{Baseline vs.\ staged agent on the synthetic benchmark ($n=21$ questions, four roles, two data sources). Data is synthetic; ground truth is computed by an independent reference implementation, not by either agent.}
\label{tab:results}
\begin{tabular}{lrr}
\toprule
\textbf{Metric} & \textbf{Baseline} & \textbf{Staged} \\
\midrule
Outcome accuracy & 0.43 & 0.95 \\
Answer correctness & 0.90 & 0.90 \\
Policy violations & 7 & 0 \\
Leakage rate & 0.33 & 0.00 \\
Confidently wrong answers & 8 & 0 \\
Refusal accuracy & 0.00 & 1.00 \\
Failure attribution & 0.33 & 1.00 \\
Latency, answer cases (ms) & 0.402 & 0.351 \\
\bottomrule
\end{tabular}
\end{table}

\subsection{Dimension 6: Failure Modes}

Document-RAG systems commonly encounter retrieval misses, context poisoning, and hallucination from irrelevant context~\cite{gao2023ragsurvey}. Structured-data agents introduce additional failure surfaces because the LLM output can determine executable data-access behavior. Table~\ref{tab:failures} summarizes five risks that are particularly salient in this setting.

\begin{table}[t]
\centering
\caption{Failure-mode taxonomy for structured-data agents. The categories are especially salient when an LLM selects schemas, resolves identifiers, and generates executable queries.}
\label{tab:failures}
\begin{tabular}{p{2.2cm}p{5.3cm}}
\toprule
\textbf{Failure Mode} & \textbf{Description} \\
\midrule
Schema hallucination & LLM generates SQL referencing nonexistent tables or columns \\
\addlinespace
Wrong joins & Tables joined on incorrect keys; structurally valid but semantically wrong results \\
\addlinespace
Identifier confusion & System uses one identifier type where another is expected (e.g., billing number vs.\ service ID) \\
\addlinespace
Auth leakage & Multi-domain response inadvertently exposes information about unauthorized domains \\
\addlinespace
Silent data subset & Query returns plausible but incomplete results due to missing join conditions \\
\bottomrule
\end{tabular}
\end{table}

\textbf{Schema hallucination} occurs when the LLM generates a query that references tables or columns not represented in the available metadata. A deterministic registry or execution-layer check can reject such references, but the check depends on current metadata.

\textbf{Wrong joins} occur when the LLM joins tables on incorrect keys, producing results that are structurally valid but semantically wrong. A representative enterprise variant is joining on an identifier column that exists in both tables but carries different meanings, such as an account number in one table and a service identifier in another.

\textbf{Identifier confusion} arises from the diversity of identifier types in enterprise systems. A billing account number, a service identifier, a contract number, and a network identifier may all refer to aspects of the same customer but are not interchangeable in queries. The LLM does not inherently understand these distinctions.

\textbf{Authorization leakage through summarization} can occur when the LLM summarizes results from multiple domains. Even if unauthorized data was excluded before summarization, the response may inadvertently reveal the excluded domain (e.g., ``I was unable to retrieve ticketing data'' reveals that ticketing data exists and was requested). Access-denial wording and response composition therefore remain part of the security boundary.

\textbf{Silent data subsets} occur when a query executes successfully but returns incomplete data, for example because a required filter, identifier mapping, or join condition was omitted. The result can look plausible and therefore requires explicit checks, domain knowledge, or comparison with a trusted reference to detect.

\subsection{Dimension 7: Latency Profile}

A structured-data agent combines LLM calls with permission lookups, entity-resolution calls, metadata search, query execution, and response compilation. Its latency is therefore shaped by both model inference and external data services, rather than by embedding lookup alone. Conditional early exits and user clarification interrupts also create response paths with different latency semantics.

\textbf{Measured observability.} A governed agent should capture stage timings and structured events for interpretation, policy decisions, entity binding, source planning, query execution, explanation, and persistence. These artifacts support latency analysis and failure attribution, but a useful report must distinguish benchmark measurements from live-traffic distributions.

\textbf{Parallelization opportunities.} Independent interpretation, metadata lookup, and policy retrieval may run concurrently. Later stages remain conditional: entity binding may pause for user input, source planning may select a direct or multi-source route, and execution may use different governed data services. The dependency graph exposes optimization points without implying a universal speedup.

\section{Reference Architecture}
\label{sec:refarch}

\begin{figure}[t]
\centering
\begin{tikzpicture}[
    node distance=0.5cm,
    block/.style={rectangle, draw, fill=blue!8, minimum height=0.6cm, minimum width=3.0cm, align=center, font=\scriptsize, rounded corners=2pt},
    authblock/.style={rectangle, draw, fill=orange!15, minimum height=0.6cm, minimum width=3.0cm, align=center, font=\scriptsize, rounded corners=2pt, thick},
    llmblock/.style={rectangle, draw, fill=green!10, minimum height=0.6cm, minimum width=3.0cm, align=center, font=\scriptsize, rounded corners=2pt},
    arr/.style={-{Stealth[length=1.5mm]}, thick},
]

\node[block] (input) {User Question};
\node[llmblock, below=0.4cm of input] (ir) {Interpret Task\\+ Scope};
\node[authblock, right=0.3cm of ir] (auth) {Policy\\Context};
\node[authblock, below=0.4cm of ir] (disamb) {Entity Binding\\+ Policy Gate};
\node[block, below=0.4cm of disamb] (schema) {Schema + Source\\Planning};
\node[block, below=0.4cm of schema] (triage) {Answerability +\\Validation};
\node[llmblock, below=0.4cm of triage] (exec) {Governed Query\\Execution};
\node[llmblock, below=0.4cm of exec] (summ) {Explain Results\\(LLM)};
\node[block, below=0.4cm of summ, fill=blue!15] (resp) {Audit +\\Provenance};

\draw[arr] (input) -- (ir);
\draw[arr] (input.east) -| (auth);
\draw[arr] (ir) -- (disamb);
\draw[arr] (auth.south) |- (disamb);
\draw[arr] (disamb) -- (schema);
\draw[arr] (schema) -- (triage);
\draw[arr] (triage) -- (exec);
\draw[arr] (exec) -- (summ);
\draw[arr] (summ) -- (resp);

\node[font=\scriptsize\itshape, text=gray, left=0.15cm of ir] {parallel};
\node[font=\scriptsize\itshape, text=gray, right=0.15cm of auth] {access};

\end{tikzpicture}
\caption{Generic reference architecture for governed structured-data agents. Green nodes may use language models, while orange nodes represent policy boundaries. Independent context lookup may run with interpretation; entity binding and policy checks constrain later source planning and execution.}
\label{fig:refarch}
\end{figure}
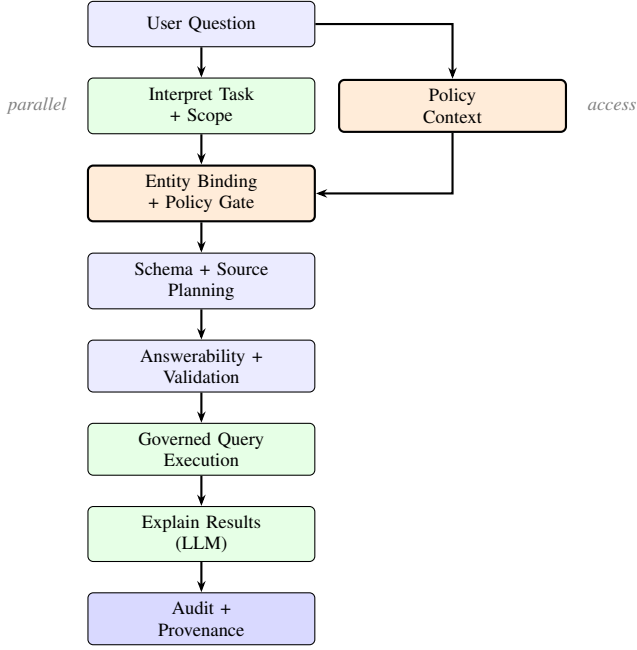

Based on the seven dimensions, we propose a reference architecture for structured-data agents. Figure~\ref{fig:refarch} shows the pipeline.

The architecture operationalizes the seven dimensions through six design principles:

\textbf{1. Schema-aware source planning.} A registry describes available sources, fields, identifiers, relationships, and execution constraints. Planning uses this metadata to narrow the sources that can answer a question.

\textbf{2. Explicit task and scope interpretation.} An interpretation component identifies the requested task, data areas, entities, and uncertainty. It may distinguish a direct lookup from a comparison or sequential investigation.

\textbf{3. Policy enforcement outside the model.} Entity, data-scope, and field-scope decisions are enforced by policy services or execution boundaries rather than by a language model's instructions. The resulting context constrains later stages.

\textbf{4. Entity binding as a control-flow step.} References in a question are bound to canonical identifiers and source-specific translations before query construction. Ambiguous matches cause clarification or abstention rather than arbitrary selection.

\textbf{5. Answerability validation before execution.} A validation stage checks whether the request has sufficient entity, temporal, source, and field context. It can reject or clarify an unsafe request before a governed query service is called.

\textbf{6. Conditional execution and provenance.} The design supports direct, multi-source, and investigative routes; records early exits; explains the returned result; and preserves enough provenance to support review and evaluation.

\textbf{Contrast with document RAG.} A minimal document-RAG pipeline can be summarized as embed, retrieve, and generate. A structured-data design adds explicit scope interpretation, policy enforcement, entity binding, source planning, validation, governed execution, explanation, and provenance. These capabilities need not be implemented as a fixed sequence: deployments can combine stages, run independent lookups concurrently, or stop when a policy or answerability condition fails.

\textbf{Hybrid systems.} Some enterprise deployments combine document RAG and structured-data querying in a single agent. For example, a user might ask ``what is the service commitment for a customer's current order?'' The commitment terms may be in a document, while order status is in a database. Hybrid systems add a routing layer that classifies each question as document-retrievable or data-queryable (or both) and dispatches to the appropriate pipeline. This routing decision is itself a form of task interpretation.

\textbf{Registry design.} A source registry is the structured-data agent's metadata foundation. It records source descriptions, fields, identifiers, relationships, policy attributes, and supported operations. Query planning consumes this metadata, while change control keeps the registry aligned with source evolution. This motivates Open Problem~2 (schema evolution).

\section{Discussion: Design Implications}
\label{sec:discussion}

A few design choices repeatedly separated safe behavior from unsafe behavior in our analysis and in the controlled study.

The most consequential is where authorization lives. If the interpretation or query-generation model is allowed to decide what a user may see, every prompt change becomes a possible security regression. Keeping policy in deterministic services or at the execution boundary, and carrying the decision as explicit state, is what let the staged agent in Section~\ref{sec:synthstudy} refuse out-of-scope requests instead of answering them. Partial authorization is the same problem in disguise. When a request spans permitted and forbidden scopes, quietly dropping the forbidden part returns a plausible but misleading answer, so the agent should reject the request, return an explained subset, or ask the user to narrow it.

The metadata gates matter more than their size suggests. Source planning and answerability validation should surface missing context and unsupported requests as explicit outcomes rather than hand every decision to a query generator. In the study this gate did most of the work: it is what made the agent abstain on unsupported metrics and ambiguous entities instead of answering confidently and wrongly.

Identifiers and concurrency fail in opposite directions. A registry that maps canonical references to source-specific identifiers keeps the query generator from guessing whether two similarly named keys are interchangeable, and this belongs in a data-integration stage rather than a prompt instruction. Concurrency is the reverse temptation. Interpretation, metadata lookup, and policy retrieval can run in parallel, but entity filtering and selection cannot be reordered ahead of the policy check without risking exposure. Parallelize the independent control-plane calls and leave the security-dependent ones in order.

\section{End-to-End Scenario Walkthrough}
\label{sec:walkthrough}

To illustrate how the seven dimensions interact, we trace a representative question through the generic reference architecture.

\textbf{Query:} A user asks: ``Which customers with overdue invoices also have unresolved support cases?''

\textbf{Step 1: Task and scope interpretation (Dims.~2--3).} The interpretation component identifies a comparison across financial and support data areas and records that an entity set, rather than one named customer, is requested.

\textbf{Step 2: Policy context (Dim.~2).} Policy services determine which data areas, entities, and fields are available to the requesting role. If the request is wholly outside scope, the system returns a denial. If only part is available, it must either narrow the request with an explanation or ask the user to revise it.

\textbf{Step 3: Entity binding (Dim.~4).} The agent determines how customer references are represented in each source. Ambiguous or unauthorized candidates are withheld from the query plan until the system can abstain or obtain clarification.

\textbf{Step 4: Source planning (Dim.~1).} A registry exposes candidate tables, fields, and join paths for the permitted data areas. The planner chooses a direct lookup, a multi-source comparison, or a sequential investigation based on the question and available relationships.

\textbf{Step 5: Answerability validation.} The system checks whether the selected sources contain the required measures, time predicates, identifiers, and join conditions. Missing context causes clarification or early termination rather than an uncontrolled query.

\textbf{Step 6: Governed execution (Dims.~1 and 6).} A validated plan is passed to an approved query service. Policy enforcement is repeated at the execution boundary so that generated query text cannot expand the authorized data set.

\textbf{Step 7: Explanation and provenance (Dims.~5 and 7).} The agent explains the result using the returned data and records the selected sources, filters, policy decisions, and unresolved limitations. The explanation must not reveal excluded data or sensitive source details.

This walkthrough is a design example, not a trace of a particular deployment. It shows why scope, policy, entity, source, validation, execution, and provenance decisions interact.

\section{Open Problems}
\label{sec:open}

Our design analysis reveals four open problems that current structured-data agent architectures do not fully solve.

\textbf{OP1: Cross-domain query composition.} When a user's question spans multiple domains (e.g., ``which customers with overdue invoices also have open tickets?''), the system must compose queries across multiple data sources and join the results. Current architectures handle each domain independently and cannot express cross-domain joins. The walkthrough above illustrates a simpler case where two domains are queried independently and results are presented side by side. True cross-domain composition---where the answer depends on joining results across domains (e.g., filtering customers who appear in both result sets)---remains architecturally unsolved without a unified query layer across data sources.

\textbf{OP2: Schema evolution.} Enterprise data sources change: columns are added, renamed, or deprecated. The schema registry must track these changes and propagate them to query generation, source planning, and execution validation. A generic architecture needs an evolution protocol that detects metadata drift, preserves compatibility where possible, and prevents stale descriptions from reaching the query-generation model.

\textbf{OP3: Explanation and provenance.} When a document-RAG system answers a question, it can cite source passages. When a structured-data agent returns a number, provenance includes the selected data source, resolved entity, filters, and execution request, but raw SQL or API payloads are not interpretable by most users. Translating this logic into a concise, verifiable query explanation remains an open evaluation problem. A natural-language ``query card'' is one possible design, but its usefulness and disclosure risks require study.

\textbf{OP4: Evaluation at enterprise scale.} Text-to-SQL benchmarks such as Spider~\cite{yu2018spider} and BIRD~\cite{li2024bird} primarily evaluate database-grounded question answering under benchmark-specific schemas. A governed structured-data benchmark would also need to represent heterogeneous sources, policy decisions, entity resolution, conditional routing, and user interruptions while avoiding proprietary data disclosure. Building and validating such a benchmark remains an open research problem.

\section{Related Work}
\label{sec:related}

\textbf{RAG systems.} Lewis et al.~\cite{lewis2020rag} introduced RAG for knowledge-intensive NLP tasks. Gao et al.~\cite{gao2023ragsurvey} provide a comprehensive survey of RAG techniques. Chen et al.~\cite{chen2024benchmarking} benchmark RAG systems across chunking, embedding, and retrieval strategies. These works focus exclusively on unstructured document retrieval.

\textbf{Text-to-SQL.} Yu et al.~\cite{yu2018spider} introduced the Spider benchmark for cross-database text-to-SQL. Li et al.~\cite{li2024bird} proposed BIRD, a benchmark emphasizing real-world database complexity. Recent LLM-based approaches~\cite{pourreza2024dinsql, gao2024texttosql} achieve strong results on these benchmarks but operate on single databases without authorization, entity resolution, or multi-domain routing.

\textbf{Enterprise data-agent architectures.} Enterprise-oriented proposals increasingly combine orchestration, data registries, metadata planning, access checks, and structured execution. A compound-AI blueprint introduces agent and data registries, task and data planners, sessions, and an execution coordinator~\cite{compoundai2025}. Analytic Agent targets governed enterprise analytics APIs and combines intent parsing, target grounding, permission validation, endpoint selection, execution, and response generation~\cite{analyticagent2026}. RUBICON argues for table-centric query processing over messy enterprise data rather than text-only integration~\cite{rubicon2026}. COGNI combines modality routing, document retrieval, and a self-correcting NL2SQL path in a conversational enterprise query engine~\cite{cogni2026}. These systems establish that enterprise data agents require more than single-shot text-to-SQL or document retrieval.

Our distinction is narrower: we organize the control dependencies among task interpretation, policy enforcement, entity binding, source planning, validation, execution, explanation, and provenance. We present these dependencies as seven architectural dimensions and analyze their relationship to a minimal document-RAG baseline.

\textbf{Agent governance.} ARBITER~\cite{arbiter2025} implements role-based access control for RAG at document granularity. Permission-Aware RAG~\cite{permaware2025} integrates IAM systems for retrieval filtering. MI9~\cite{mi9_2025} proposes runtime governance for agentic AI with agency-risk indexing. Our focus is the placement and interaction of access checks within a structured-data query workflow rather than document filtering alone.

\textbf{Bounded autonomy.} The typed action contracts architecture~\cite{bounded2025} constrains LLM agents to pre-defined action schemas with validation before execution. This shares our principle of schema validation before query execution, but focuses on general enterprise actions rather than structured-data querying specifically.

\textbf{Multi-agent evaluation.} MAESTRO~\cite{maestro2025} provides a framework-agnostic evaluation suite for multi-agent systems, focusing on execution traceability and system-level comparison across frameworks. MASEval~\cite{maseval2025} demonstrates that framework choice impacts performance as much as model choice. TraceElephant~\cite{traceelephant2026} introduces failure attribution benchmarks using full execution traces. These works evaluate multi-agent orchestration but do not address the structured-data-specific dimensions we identify (entity resolution, schema hallucination, multi-layer authorization).

\textbf{SQL safety and governance.} Recent work on LLM SQL safety includes trust scoring for hallucinated tables and columns~\cite{sqltrust2025} and intent-based database access protocols that replace raw SQL generation with structured intent objects~\cite{mdbp2025}. These approaches address schema hallucination at the query level but do not consider the broader pipeline requirements (entity resolution, multi-domain routing, authorization) that structured-data agents face.

\section{Threats to Validity}
\label{sec:threats}

\textbf{Conceptual scope.} The paper presents a reference architecture rather than a complete implementation. The seven dimensions may not be exhaustive, and other deployments may expose additional requirements. We limit the claims to design responsibilities and control dependencies that can be evaluated in a concrete system.

\textbf{Enterprise specificity.} The authorization and entity-resolution patterns are shaped by requirements such as persona-based access, entity-level restrictions, heterogeneous identifiers, and governed data services. Less constrained personal analytics tools may not require all seven dimensions. The framework should therefore be read as a set of requirements for governed enterprise agents, not as a universal design for every structured-data interface.

\textbf{Synthetic rather than deployment evaluation.} The empirical comparison in Section~\ref{sec:synthstudy} uses a small synthetic benchmark under controlled conditions. It does not report production incident rates, live-traffic latency, or results on a particular organization's data. The synthetic results show that the architectural stages have a measurable effect, but claims about enterprise-scale effectiveness remain to be validated on disclosed or deployed systems.

\textbf{Transferability.} The examples and terminology are intentionally generic, which improves portability but removes integration details needed for exact reproduction. The intended contribution is a set of architectural responsibilities, evaluation measures, and control dependencies that other researchers can instantiate and test.

\section{Conclusion}
\label{sec:conclusion}

The shift from document retrieval to structured-data querying exposes architectural requirements across seven dimensions: retrieval semantics, authorization granularity, intent recognition, entity resolution, evaluation, failure modes, and latency profiles. A minimal document-RAG baseline does not make these requirements explicit because it assumes that the principal operation is selecting text context.

We presented a generic reference architecture for governed structured-data agents: explicit task and scope interpretation, policy enforcement, entity binding, schema-aware source planning, answerability validation, governed execution, explanation, and provenance. We also organized five risks that become especially important when model outputs influence structured-data access: schema hallucination, wrong joins, identifier confusion, authorization leakage, and silent data subsets.

A controlled synthetic study showed that adding these stages to a direct translate-and-execute baseline removed the baseline's authorization violations and raised outcome accuracy from 0.43 to 0.95 under controlled conditions, which suggests the extra stages do real work rather than adding overhead. Four open problems remain: cross-domain query composition, schema evolution management, query provenance for non-technical users, and enterprise-scale evaluation. Rather than proposing a new language model or query algorithm, this paper contributes a design framework, and a first empirical test of it, that makes the control-flow, security, and evaluation requirements of governed structured-data agents explicit and testable.

\bibliographystyle{IEEEtran}

\end{document}